\RequirePackage{fix-cm}
\documentclass{svjour3}                     
\smartqed  
\usepackage{graphicx}
\usepackage{amsmath}

\usepackage{latexsym}
\newsavebox{\astrutbox}
\sbox{\astrutbox}{\rule[-5pt]{0pt}{20pt}}

\def\der#1#2{{\partial #1\over \partial #2}}

\def\be{\begin{equation}}
\def\ee{\end{equation}}
\def\bea{\begin{eqnarray}}
\def\eea{\end{eqnarray}}
\def\bse{\begin{subequations}}
\def\ese{\end{subequations}}
\def\bsea{\begin{subeqnarray}}
\def\esea{\end{subeqnarray}}
\def\({\left (}
\def\){\right )}
\def\[{\left [}
\def\]{\right ]}
\def\<{\left <}
\def\>{\right >}

\begin{document}

\title{Toward a thermo-hydrodynamic like description of Schr\"{o}dinger equation via the Madelung formulation and Fisher information}
\titlerunning{}

\author{Eyal Heifetz, Eliahu Cohen}


\institute{Eyal Heifetz \at Department of Geosciences, Tel-Aviv University, Tel-Aviv, Israel\\
  {\email eyalh@post.tau.ac.il} \and Eliahu Cohen \at School of Physics and Astronomy, Tel-Aviv University, Tel-Aviv, Israel\\
  {\email eliahuco@post.tau.ac.il}}


\date{\today}

\maketitle

\begin{abstract}

We revisit the analogy suggested by Madelung between a non-relativistic time-dependent quantum particle, to a fluid system which is pseudo-barotropic, irrotational and inviscid. We first discuss the hydrodynamical properties of the Madelung description in general, and extract a pressure like term from the Bohm potential. We show that the existence of a pressure gradient force in the fluid description does not violate Ehrenfest's theorem since its expectation value is zero. We also point out that incompressibility of the fluid implies conservation of density along a fluid parcel trajectory and in $1D$ this immediately results in the non-spreading property of wave packets, as the sum of Bohm potential and an exterior potential must be either constant or linear in space.

Next we relate to the hydrodynamic description a thermodynamic counterpart, taking the classical behavior of an adiabatic barotopric flow as a reference. We show that while the Bohm potential is not a positive definite quantity, as is expected from internal energy, its expectation value is proportional to the Fisher information whose integrand is positive definite. Moreover, this integrand is exactly equal to half of the square of the imaginary part of the momentum, as the integrand of the kinetic energy is equal to half of the square of the real part of the momentum. This suggests a relation between the Fisher information and the thermodynamic like internal energy of the Madelung fluid. Furthermore, it provides a physical linkage between the inverse of the Fisher information and the measure of disorder in quantum systems - in spontaneous adiabatic gas expansion the amount of disorder increases while the internal energy decreases.

\end{abstract}

\section{Introduction}

A year after Erwin Schr\"{o}dinger published his celebrated equation, Erwin Madelung showed (in 1927) that it can be written in a hydrodynamic form \cite{Mad}. Madelung's representation has a seemingly major disadvantage by transforming the single linear Schr\"{o}dinger equation into two nonlinear ones. Nonetheless, despite of its additional complexity, the hydrodynamic analogy provides important insights with regard to the Schr\"{o}dinger equation
\cite{Tak,Sch,Son,Tse1,Tse2}. The Madelung equations (ME) describe a compressible fluid, and compressibility yields a linkage between hydrodynamic and thermodynamic effects. The work done by the pressure gradient force to expand the flow transforms internal thermal microscopic kinetic energy to the macroscopic hydrodynamic kinetic energy of the flow. In this paper we wish to examine to what extent such a linkage can be made and what added value it provides in understanding quantum systems.




ME can be obtained when taking the non-relativistic time dependent Schr\"{o}dinger equation (TDSE) of a particle with mass $m$, under the presence of an external potential $U({\bf r},t)$:
\be
i\hbar\der{\Psi}{t} = \hat{H}\Psi = \({\hat{p}^2 \over 2m} +U\)\Psi =
\(-{\hbar^2\over 2m}\nabla^2 +U\)\Psi.
\ee
Assuming that the wave function $\Psi$ is continuous and can be written in the polar form
\be
\Psi({\bf r},t) = \sqrt{\rho}({\bf r},t)e^{iS({\bf r},t)/\hbar},
\ee
then together with the de Broglie guiding equation for the velocity
\be
{\bf u} = \nabla{\tilde S}
\ee
(where the tilde superscript represents hereafter a quantity per unit mass $m$, so that ${\tilde S}  = {S\over m}$), the real part of the TDSE becomes the continuity equation
\be
{D\over Dt}\ln\rho = - \nabla\cdot{\bf u}\, ,
\ee
and the imaginary part becomes
\be
\der{\tilde S}{t} = -({\tilde K} + {\tilde Q} + {\tilde U}),
\ee
where ${\tilde K} = {\bf u}^2/2$ is the kinetic energy per unit mass and
\be
\tilde{Q}  = - {\hbar^2 \over 2m^2}{\nabla^2 \sqrt{\rho} \over \sqrt{\rho}}
\ee
is the Bohm potential per unit mass \cite{Bohm}.
Generally $({\bf u}\cdot\nabla){\bf u} = \nabla{\tilde K} + {\boldsymbol \omega}\times{\bf u}$, where ${\boldsymbol \omega} = \nabla\times{\bf u}$ is the vorticity,
however for a potential flow in the form of (3), the flow is irrotational, i.e.,  ${\bf \omega} = 0$, and therefore when applying the nabla operator on (5)
we obtain
\be
{D\over Dt}{\bf u} = -\nabla \tilde{Q}(\rho) -\nabla{\tilde U},
\ee
where ${D\over Dt} \equiv \der{}{t} + {\bf u}\cdot\nabla$, is the material (Lagrangian) time derivative of a fluid element along its trajectory.
Equations (4) and (7) map the TDSE to a pseudo-barotropic, inviscid flow where (7) is its inviscid Navier-Stokes (i.e., Euler) equation.
As indicated by \cite{Wal}, in order for this formalism to be equivalent to the Schr\"{o}dinger equation, one should also add a quantization condition.

The paper is organized as follows. In Section 2 we examine the hydrodynamical properties of the Madelung equations. In Section 3 we relate them to their thermodynamical properties, and show how the Fisher information provides an analogy to the internal thermal energy of the flow. Discussion of the results appears in Section 4.

\section{Hydrodynamic properties of the Madelung equations}

\subsection{\bf The continuity equation}

It is somewhat surprising that we can describe the quantum state of a single particle in terms of a fluid whose mass density $\rho_{fluid}({\bf r},t) = \delta M/\delta V$  is the probability density $\rho({\bf r},t)$ of the wave function  to find the particle $m$ in location ${\bf r}$ at time $t$
($\delta M$ is an infinitesimal ``fluid mass'' occupying an infinitesimal volume $\delta V$, where $M$ is non-dimensional since it represents probability).
The continuity equation (4) is usually represented in its Eulerian form
\be
\der{\rho}{t} = -\nabla\cdot{\bf J},
\ee
where ${\bf J}=\rho{\bf u}$ is the flow mass flux. In its Lagrangian representation however, (4) simply reflects the statement that a ``fluid parcel'' conserves its mass $\delta M$ as it moves with velocity ${\bf u}$, i.e., ${D\over Dt}(\delta M) = 0$. (4) is then obtained when noting that $\nabla\cdot{\bf u} = {D\over Dt}\ln (\delta V)$ is the flow compressibility term. Hence, when the flow is incompressible ${D\over Dt}\rho = 0$, and that is to say that the density is conserved along a ``fluid parcel'' trajectory.
This is indeed the case when the wave function is represented by a single plane wave since then ${\bf u} = \hbar{ \bf \tilde k}$, where ${\bf k}$ is the wavenumber vector. Nonetheless, interference between two plane waves or more yields $\nabla\cdot {\bf u} \neq 0$ in general, and thus the probability to find a particle along a trajectory constructed from (3) varies along the trajectory itself. In $1D$, incompressibility implies that ${\bf u}$ is not a function of space, and therefore $\rho$ is non-spreading whether or not the flow is accelerating.


\subsection{\bf Relating the Bohm potential to a pseudo barotropic pressure}

For an inviscid flow, in the presence of an exterior potential $U$, the Euler equation (Newton's second law essentially) reads:
\be
{D\over Dt}{\bf u} = -{1\over \rho}\nabla P -\nabla{\tilde U},
\ee
where the pressure $P$ is a known thermodynamic property.
Furthermore, if the flow is barotropic, that is the pressure $P$ is a function of density only, the pressure gradient force (PGF) can be written as a perfect gradient
\be
-{1\over \rho}\nabla P(\rho) = - \nabla {\tilde Q(\rho)},
\ee
where ${\tilde Q} = \int{dP \over \rho}$ is defined up to an unspecified time dependent gauge function.
Hence, taking then the curl of (9) we obtain
\be
\der{\boldsymbol \omega}{t} = \nabla\times({\bf u}\times{\boldsymbol \omega}),
\ee
therefore, if ${\boldsymbol \omega}(t=0) = 0$, the vorticity remains zero at all times and thus the velocity field can be represented as a potential flow of the form of (3). In regions where $\rho = 0$, the vorticity can be non-zero in principle, as in superfluid singularities, however we refrain here from discussing such cases.

It is worth noting that Euler had derived equation (9) in 1757 more than a century before the thermodynamic kinetic theory of gases was established by Maxwell and Boltzamnn in 1871. Hence, the knowledge that the pressure results from the aggregated effect of microscopic random motions of molecules or atoms impacting each other, was not necessary to Euler to formulate it as a macroscopic force per unit area that is somehow associated with the intrinsic properties of the fluid. In a sense, this is what we try to do next with respect to the Madelung equation (7). Here, the starting point is the known function of the Bohm potential $Q(\rho)$, hence we can solve (10) for $P$ to obtain:
\be
P = \Pi(\rho) +f(t)
\ee
where
\be
\Pi(\rho) = -\({\hbar\over 2m}\)^2\rho\nabla^2\ln\rho,
\ee
and $f(t)$ is a time dependent gauge function. Here we used the identity
\be
{\nabla^2\alpha\over \alpha} = \nabla^2\ln\alpha +(\nabla\ln\alpha)^2,
\ee
to extract $\Pi$ from ${\tilde Q}$ in (10).
As $\Pi$ depends on spatial derivatives of the density, it cannot be defined as a proper thermodynamic pressure, hence a possible appropriate name for it could be a pseudo pressure or a pressure like term \footnote{Indeed, more complex interpretations have been suggested with regards to the quantum pressure \cite{Sch,Tse1}, however the proposed pressure like term still seems to us the most straightforward one as we are seeking for the simplest analogy. A second rank tensor of the Cauchy stress tensor, embedding shear terms, as well as the introduction of turbulence to the ME, or the representation of it in terms of nonlinear diffusion are interesting interpretations, however they involve an amount of complexity that we try to avoid when considering the dynamics of a single quantum particle.}. An example that makes sense of (13) is a normal probability density function, as it is straightforward to see that $\Pi \propto \rho$, which is true for an ideal isothermal perfect gas.


In the absence of an external potential $U$, it is surprising that the quantum state of a free particle is mapped into a fluid that can be accelerated by a force, even if this force is the PGF which is internal to the flow. However, defining the acceleration as ${\bf a} = {D{\bf u}\over Dt}$, its expectation value can be found to be
\be
\<{\bf a}\> = \int \Psi^{*} {\bf a} \Psi dV = \int \rho {\bf a}\, dV   = -\int P d{\bf A}
\ee
(where the integrations are taken over the whole domain $V$).
Thus, if $\rho$ is bounded within the domain and vanishes at the domain boundary $A$, then the net expectation value of the acceleration is zero and Ehrenfest's theorem is not violated.

\subsection{\bf Flow incompressibility and non-spreading wave packets}

As pointed out in Section (2.1), when the flow is incompressible $(\nabla\cdot{\bf u}=0)$, density is conserved along ``fluid parcel'' trajectories.
It is worth noting that this statement is equivalent to the one that a wave packet following such trajectory is non-spreading.
Furthermore, since ${D{\bf u}\over Dt} = \der{{\bf u}}{t} +{\boldsymbol \omega}\times{\bf u} + \nabla({\bf u}^2/2)$, and the flow is irrotational, (7) becomes
\be
\der{{\bf u}}{t} = -\nabla \({\tilde K} + \tilde{Q} +{\tilde U}\).
\ee
Specifically, for $1D$ flow, incompressibility $(\der{u}{x}=0)$ implies that $u$ can be only a function of time and (16) becomes
\be
\der{u}{t} = -\der{}{x} \({\tilde Q} +{\tilde U}\) = g(t),
\ee
posing the constraint on the potentials: ${\tilde Q} +{\tilde U} = a(t)x +b(t)$. Among the potentials mentioned in the literature which satisfy this non spreading condition are:
$i)$ the free Airy wave packet  ($\sqrt\rho  \propto Ai(x)$ \& $U = 0$) \cite{Ber}; $ii)$ the gravitational ``quantum  bouncer'' ($\sqrt\rho \propto Ai(x)$ \&
$U \propto x$) \cite{LL}; $iii)$ the ground state of the harmonic oscillator ($\sqrt\rho \propto e^{-x^2}$ \& $U  \propto x^2$) \cite{LL}.

\section{\bf Thermodynamic like properties of the Madelung equations}

In this section we examine to what extent one may associate thermodynamic like properties with the Madelung description, in reference to the
thermodynamic of classical barotorpic conservative (adiabatic) flows.

\subsection{\bf Energy conservation}

Consider a classical, adiabatic (entropy conserving, hence inviscid), barotopric flow. In the absence of heat exchange, the first law of thermodynamics reads,
$dI = -PdV$, where $I$ is the  thermal internal energy, representing the macroscopic aggregated effect of the microscopic random thermal fluctuations. Hence, during an adiabatic process, compression of a fluid parcel by its surroundings performs work that increases its internal energy. If, however, the flow is incompressible, the internal energy remains unchanged. When following materially a fluid parcel in motion, the adiabatic first law is transformed into
\be
\rho {D\over Dt}{\tilde I} = -P{D\over Dt}\ln (\delta V) = -P\nabla\cdot {\bf u}.
\ee
Multiplying (9) by $\rho{\bf u}$ and combining it with (18), then for a time independent external potential $U$, we obtain
\be
\rho{D\over Dt}\({\tilde K} + {\tilde I} + {\tilde U}\) = -\nabla\cdot({\bf u}P).
\ee
Thus, the total energy of a fluid parcel per unit mass
\be
{\tilde E_{Cl}} = ({\tilde K} + {\tilde I} + {\tilde U}),
\ee
is not materially conserved simply because the fluid parcel is not an isolated system (the subscript Cl denotes classical fluid).
The surrounding pressure can change the parcel internal energy by compression and the pressure gradient can accelerate the parcel and hence change its kinetic energy.
Nonetheless, the overall total energy of the fluid is conserved in the domain averaged sense.
We can use (4) to obtain that for any scalar field $\alpha$, $\rho{D\alpha\over Dt} = \der{}{t}(\rho \alpha)  + \nabla\cdot(\rho{\bf u}\alpha)$, and therefore if all fluxes vanish on the domain boundaries we obtain
\be
\der{}{t}\< {\tilde E_{Cl}}\> =\der{}{t}\int \rho {\tilde E_{Cl}}dV = 0.
\ee

On the other hand, direct calculation of the energy expectation value from the Schr\"{o}dinger equation (1), yields the
conservation of the total energy (assuming as well that all fluxes vanish on the domain boundaries):
\be
\<{\tilde E_{Qu}}\> = {1\over m}\int\Psi^{*}\(-{\hbar^2\over 2m}\nabla^2 +U\)\Psi dV=
\<{\tilde K} + {\tilde Q}+ {\tilde U}\>,
\ee
where ${\tilde Q}$ is the Bohm potential (per unit mass) given by (6), and the subscript Qu refers to a quantum like fluid. Comparing (20) with (22) suggests therefore that $\<{\tilde Q}\> = \<{\tilde I}\>$. A similar argument, arising from a different perspective, has been recently suggested in \cite{Dennis}.
The Bohm potential by itself, however, is not a positive-definite term, as is expected from internal energy. Nonetheless its expectation value yields, after integration by parts,
\be
\<{\tilde Q}\> = \< {1\over 2}\[{\hbar\over 2m}\nabla(\ln\rho)\]^2 \> = {1\over 2}\({\hbar\over 2m}\)^2 FI,
\ee
where $FI = \int{1\over \rho}(\nabla \rho)^2 dV = \int \rho[\nabla(\ln\rho)]^2 dV$ is the Fisher information, in the form presented in \cite{Reg,Fri}.
For instance for normal distribution the Fisher information is inversely proportion to its variance, and according to the Cramer-Rao bound the general variance of any unbiased estimator (like the location of a particle ${\bf r}$ at time $t$) is always larger (or equal for normal distribution) to $1/FI$  \cite{Fri}. Therefore we can choose to define the positive definite internal energy per unit mass as
\be
{\tilde I} = {1\over 2}\[{\hbar\over 2m}\nabla(\ln\rho)\]^2.
\ee
Furthermore, this choice suggests a relation between the quantum thermal energy and the imaginary part of the momentum field.
Writing the momentum in the form of
\be
\hat{\bf p}\,\Psi =[-i\hbar\nabla\ln\Psi]\Psi \equiv {\bf p}\,\Psi,
\ee
we can define the complex velocity field as
\be
{\bf v} \equiv {\tilde {\bf p}} = {\bf v}_r +i{\bf v}_i
\ee
where
\be
{\bf v}_r = {\bf u} = \nabla {\tilde S}, \hspace{1cm}
{\bf v}_i = -{\hbar \over 2m}\nabla(\ln\rho).
\ee
This partition has been suggested, by \cite{Rec,Esp} in different contexts.
Imaginary momentum is a somewhat strange concept, however here we simply note that in the same sense that under the de Broglie guiding equation (3), the macroscopic kinetic energy of the flow is ${\tilde K} = {1\over 2}{\bf v}_r^2$, under (24) ${\tilde I} = {1\over 2}{\bf v}_i^2$, suggesting that within this analogy, the microscopic thermal motion is represented by the imaginary momentum. Moreover, the expectation value of random fluctuations
must be zero by definition, and indeed the expectation value of each component of ${\bf v}_i$ is proportional to the Fisher score which is zero.

As pointed out by \cite{Reg,Fri} the inverse of the Fisher information can be used to measure the degree of disorder in the system and it tends to increase with time. In this sense, the Fisher information measures the ``narrowness'' of the position distribution around the actual position of the particle \cite{Reg}.
The relation of the internal energy of the Madelung flow to the Fisher information provides a physical interpretation of this statement.
Consider a gas expanding spontaneously in an adiabatic process. The degree of disorder increases and according to the first law of thermodynamics (18) the internal energy decreases. The fact that the Fisher information may provide better estimation to disorder than the negative entropy (negentropy), may result from the fact that during adiabatic expansion the thermodynamic entropy remains unchanged.
Furthermore, in the absence of an external potential, $\<{\tilde K} + {\tilde I}\>$ is conserved and if $\<{\tilde I}\>$ decreases during expansion the kinetic energy and momentum become more pronounced. This may suggest a different angle to the relation between the Fisher information and the Heisenberg uncertainty principles \cite{FIuncer}.



\subsection{\bf The Bohm potential and pseudo enthalpy}

Equation (10) can be rewritten in the form
\be
\nabla {\tilde Q} =  {1\over \rho}\nabla P = \({P\over \rho^2}\)\nabla\rho + \nabla\({P\over \rho}\),
\ee
and for  classical barotropic adiabatic flows we can substitute the continuity equation (4) with the first law of thermodynamics (18) to obtain
\be
{D\over Dt}{\tilde I} = {P\over \rho^2} {D\over Dt}\rho   \hspace{0.25cm} \Rightarrow \hspace{0.25cm}
{d\over d\rho}{\tilde I}_{Cl}(\rho) = {P_{Cl}(\rho)\over \rho^2}.
\ee
Hence for classical barotropic adaibatic flows, (29) implies that
\be
\[{\tilde Q} = \({\tilde I} + {P\over \rho}\) +{\tilde Be}(t) = {\tilde Ent} +{\tilde Be}(t)\]_{Cl},
\ee
where ${\tilde Ent} = \({\tilde I} + {P\over \rho}\)$ is the enthalpy per unit mass and ${\tilde Be}(t)$ is a time dependent gauge function that is equal to the Bernoulli potential (as will be seen in the next subsection). On the other hand, for the quantum flow, (6), (13) and (24) indicate that the Bohm potential satisfies
\be
\[{\tilde Q} = \(-{\tilde I} + {\Pi\over \rho}\)\]_{Qu}.
\ee
This modified form of quantum enthalpy does not violate the equality $\<{\tilde Q}\> = \<{\tilde I}\>$, since it is straightforward to show that
$\int\Pi dV = 2\<{\tilde I}\>$. Hence, although $\Pi(\rho)$ is not a positive definite quantity, its domain averaged value is always positive.

\subsection{\bf The Bernoulli and Hamilton-Jacobi equations}

For completeness we discuss the relation between the Bernoulli and Hamilton-Jacobi equations in classical and quantum systems.
If the flow is potential of the form of (3) we can obtain from (7) the time dependent Bernoulli equation
\be
\nabla {\tilde Be}(t) = 0, \hspace{0.25cm}  {\tilde Be}(t) =  \der{\tilde S}{t} + ({\tilde K} + {\tilde Q} + {\tilde U}),
\ee
so that together with (30)
\be
\[\der{\tilde S}{t} = -\({\tilde K} + {\tilde Ent} + {\tilde U}\) = -{\tilde H}\]_{Cl},
\ee
where ${\tilde H}_{Cl}$ differs from ${\tilde E}_{Cl}$ in (20) by the term of ${P\over \rho}$. Moreover, since the thermodynamic pressure is positive definite it is clear that $\[\<{\tilde H}\> \neq \<{\tilde E}\>\]_{Cl}$. With this alert in mind, (34) can be regarded as the fluid mechanics version of the Hamilton-Jacobi equation. Since the velocity potential is only a function of time and space, ${\tilde S}={\tilde S}({\bf r},t)$ (and not of the momentum ${\bf p}$), the material derivative ${D{\tilde S}\over Dt} = \der{\tilde S}{t} +2{\tilde K}$, is the total time derivative in the phase space of $({\bf r}, {\bf p})$. Substituting in (33) we obtain
\be
{D{\tilde S}\over Dt} = {\tilde K} -\({\tilde Ent} + {\tilde U}\) = {\tilde L}   \hspace{0.25cm} \Rightarrow \hspace{0.25cm}
{\tilde S} = \int {\tilde L} Dt,
\ee
where ${\tilde L}$ and the velocity potential ${\tilde S}$ may serve respectively, as the Lagrangian and action of this system (where the material integration in time follows the fluid parcel).
Requiring the action to be stationary by Hamilton's principle, we obtain that (7) becomes the Euler-Lagrange set of equations.
To the best of our knowledge this relation between the Bernoulli and the Hamilton-Jacobi equations has not been acknowledged in the literature.

In the quantum realm the starting point is equation (5), thus ${\tilde Be}(t)$=0 and
${\tilde H}_{Qu}= ({\tilde K} -{\tilde I} + {\Pi \over \rho} + {\tilde U})$. But unlike the classical counterpart, here indeed
$\[\<{\tilde H}\> = \<{\tilde E}\>\]_{Qu}$.

\section{Discussion}

The Madelung equations suggests an alternative appealing description to the Schr\"{o}dinger equation by associating the behavior of a non relativistic quantum particle to the dynamic of a fluid. This provides an additional important intuition, although that in principle the nonlinear fluid dynamic equations are much more complex than the single linear Schr\"{o}dinger one. Any compressible flow exchanges kinetic energy with thermal internal energy as both result from the motion of the particles composing the flow. The Madelung flow is compressible, however the Madelung equations do not provide an explicit description of the involved thermodynamical processes. Moreover, the Madelung flow is a (pseudo) barotropic, potential (and hence irrotational), inviscid flow. The thermodynamics of classical flows which encompasses such properties, is simple, elegant and intuitive. Therefore, our motivation here was to examine to what extent we can relate to the Madelung flow, thermodynamic like properties. Throughout the text we repeat using terms such as ``like'' and ``pseudo'', in order to keep in mind that the Madelung fluid is a pure analogy. As pointed out in section 2.1 the  ``fluid density'' is the probability density function to find a single quantum particle in location ${\bf r}$ at time $t$.

In the first part of this paper we revisited the Madelung hydrodynamic analogy to the  Schr\"{o}dinger description in the presence of an exterior potential. The aim was to present this alternative, and somewhat surprising view, in a way that is appealing both to quantum physicists and classical fluid dynamicists. Furthermore, a few examples have been briefly discussed (with respect to 1D non spreading wave packets) in order to exemplify how the  hydrodynamic perspective can be intuitive and helpful to understand the behavior of quantum systems.

In the second part we suggested a complementary, thermodynamic description to the Madelung equations, where the classical thermodynamics is taken as a reference.
We find that conservation of energy, implied from the Schr\"{o}dinger equation, suggests that the expectation value of the Bohm potential is proportional to the expectation value of the internal energy in classical fluids. However, the Bohm potential is not a a positive definite function of density, as is expected from
the internal energy of barotropic fluids. Nevertheless, since the expectation value of the Bohm potential is proportional to the Fisher information, and the latter's integrand is a positive definite function of density, we related it to an equivalent thermal internal energy. This provides a direct physical link between the decrease in Fisher information and the amount of disorder increase spontaneously with time. According to the Cramer-Rao bound the variance of the Fisher information is at least inversely proportional to the variance of the probability density function of the particle location, and as disorder increases with time the Fisher information decreases and the probability variance increases. In the compressible adiabatic fluid analogy, spontaneous adiabatic expansion of fluids decreases their internal energy and increases the amount of disorder in the fluids. Since in adiabatic processes the thermodynamic entropy remains constant by definition, this analogy may strengthen the claims that the (inverse of the) Fisher information is a more suitable measure of disorder in quantum systems than entropy.

Moreover, the quantum momentum operator is generally a complex function, however the de Broglie guiding equation makes use only of its real component. Nonetheless, if the amplitude of the density is space dependent the momentum has an imaginary counterpart. We find it elegant that as the macroscopic kinetic energy of the quantum fluid is proportional to the square of the real part of the momentum, the internal energy is proportional to the square of the imaginary one. Furthermore, the expectation value of components of the imaginary part of the momentum is zero (proportional to the Fisher score), which is expected for random thermal fluctuations. This suggests that within the fluid analogy the real part of the momentum represents the domain averaged momentum of the fluid and the fluctuation from this mean is represented by the imaginary part.

The classical thermo-hydrodynamics of a barotropic conservative fluid cannot be mapped however, as is, to the Madelung fluid. This results from the different starting points of the two systems. In the classical form the pressure is a well defined thermodynamic property that is a function of density itself (and not of spatial gradients of density). Then the momentum dynamics is described by the Euler equation where the pressure gradient force can be introduced as a
perfect gradient of a function that is density dependent but can be defined up to some additional time dependent gauge function. Furthermore, an internal thermal energy exists and obeys the first law of thermodynamics and this implies that the density dependent function of the perfect gradient is the enthalpy of the system.
Domain energy conservation is obtained when transforming the momentum equation into mechanical energy one and augmenting it with the first law of thermodynamics and the continuity equation. One can also obtain the Bernoulli (which is the Hamilton-Jacobi) equation from the Euler equation up to the undefined Bernoulli potential time dependent function.

In the quantum case, the real part of the Schr\"{o}dinger equation is mapped into the classical continuity equation, however its imaginary part is mapped into a
Bernoulli like equation with zero gauge function and a well defined density dependent Bohm potential which is not the Enthalpy of the system. Then the Euler equation for barotorpic flow is obtained when the nabla operator is applied on the Bernoulli like equation. Hence, although the Euler equation in the two systems looks the same, it is obtained in two fundamental different ways. We can then extract a pressure like form from the Bohm potential but this pseudo barotropic pressure has a function that depends on the density gradient rather then the density itself, and moreover can be defined only up to a time dependent gauge function. Furthermore, no internal energy is defined a-priori from thermodynamic considerations, and the first law of thermodynamics is not a postulate of the system. On the other hand, the domain averaged conservation of the total energy is obtained directly from the Hamiltonian of the Schr\"{o}dinger equation. Although it looks like the Bohm potential plays the role of the internal energy in the Hamiltonian it is not a positive definite quantity, as opposed to the Fisher integrand that provides the same expectation value as the Bohm potential.


Overall, the suggested thermodynamic description of the Madelung flow is formed here using a top-down approach in the sense that we do not build the macroscopic thermodynamic parameters bottom-up from a comprehensive statistical microscopic theory. This is somewhat similar to the approach taken by Euler when deriving the simplified momentum equation of fluid flows before the thermodynamic kinetic theory of gases has been developed.

In future works it could be interesting to relate this suggested formalism to Zitterbewegung in a similar manner to \cite{Rec} and also to the zero point field in QED \cite{Rueda}. \\ \\

%

{\bf Acknowledgements}\\
The authors are grateful to Sonja Rosenlund Ahl for sharing her original thoughts. E.H. is grateful to Rachel Heifetz and Nili Harnik for their constructive comments.
E.C. was supported in part by the Israel Science Foundation Grant No. 1311/14

%

\end{document}